\documentclass[9pt,shortpaper,twoside,web]{ieeecolor}
\usepackage{generic}
\usepackage{cite}
\usepackage{amsmath,amssymb,amsfonts}
\usepackage{algorithmic}
\usepackage{graphicx}
\usepackage{textcomp}
\usepackage[table,xcdraw]{xcolor}
\usepackage{multirow}
\usepackage{float}
\usepackage{subfigure}
\usepackage{soul}

\def\BibTeX{{\rm B\kern-.05em{\sc i\kern-.025em b}\kern-.08em
 T\kern-.1667em\lower.7ex\hbox{E}\kern-.125emX}}
\begin{document}
\title{Discriminative Kernel Convolution Network for Multi-Label Ophthalmic Disease Detection on Imbalanced Fundus Image Dataset}
\author{Amit Bhati, Neha Gour, Pritee Khanna and Aparajita Ojha

\thanks{Amit Bhati, Neha Gour, Pritee Khanna, and Aparajita Ojha are with the Computer Science and Engineering Discipline, PDPM Indian Institute of Information Technology, Design and Manufacturing, Jabalpur, India (e-mail: pkhanna@iiitdmj.ac.in). }}

\maketitle

\begin{abstract}
It is feasible to recognize the presence and seriousness of eye disease by investigating the progressions in retinal biological structure. Fundus examination is a diagnostic procedure to examine the biological structure and anomaly of the eye. Ophthalmic diseases like glaucoma, diabetic retinopathy, and cataract are the main reason for visual impairment around the world. Ocular Disease Intelligent Recognition (ODIR-5K) is a benchmark structured fundus image dataset utilized by researchers for multi-label multi-disease classification of fundus images. This work presents a discriminative kernel convolution network (DKCNet), which explores discriminative region-wise features without adding extra computational cost. DKCNet is composed of an attention block followed by a squeeze and excitation (SE) block. The attention block takes features from the backbone network and generates discriminative feature attention maps. The SE block takes the discriminative feature maps and improves channel interdependencies. Better performance of DKCNet is observed with InceptionResnet backbone network for multi-label classification of ODIR-5K fundus images with 96.08 AUC, 94.28 F1-score and 0.81 kappa score. The proposed method splits the common target label for an eye pair based on the diagnostic keyword. Based on these labels oversampling and undersampling is done to resolve class imbalance.  To check the biasness of proposed model towards training data, the model trained on ODIR dataset is tested on three publicly available benchmark datasets. It is found to give good performance on completely unseen fundus images also.
\end{abstract}


\section{Introduction}
\label{sec:introduction}
\IEEEPARstart{O}phthalmic diseases are leading cause of blindness worldwide. A report published by the world health organization (WHO) in 2021 says that around 2.2 billion people are visually impaired, and almost half of these are preventable based on timely detection and treatment \cite{ref_32}. The human retina is a light-sensitive layer of tissues in the rear end of the eye. The incident light is converted into neural signals through the receptors on retina and handled by the brain's visual cortex to generate a picture. The retina gets affected by different abnormalities which influences vision \cite{ref_1}. Fundus, fluorescein angiography, and optical coherence tomography (OCT) are standard modalities used by experts to investigate ophthalmic diseases \cite{ref_2}. Fundus imaging is the primary image modality utilized for clinical examination of ophthalmic diseases.

\begin{figure}[!t]
\centering
\includegraphics[width=1.0in]{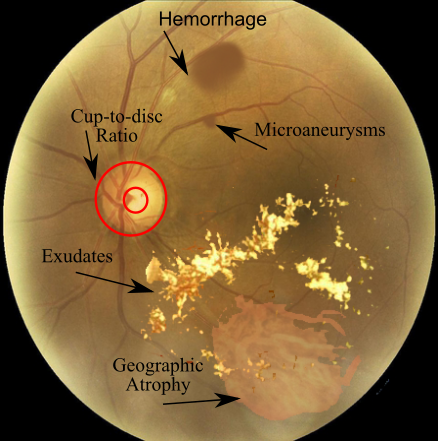}
\caption{Fundus image of an eye with different kinds of abnormalities \cite{ref_31}.}
\label{fig1}
\end{figure}

The presence of ophthalmic diseases can be recognized by observing abnormalities close to different retinal areas like optic nerve, veins, macula, optic plate, etc. Early identification of retinal abnormalities depicted in Fig \ref{fig1} is crucial, yet difficult, as a few signs appear in the beginning stage. Deep Neural Networks (DNN) are effectively utilized for retinal vessel segmentation, lesion detection \cite{ref_7, ref_8}, and glaucoma or diabetic retinopathy stage classification \cite{ref_9, ref_10}. Traditional single-label classification, also known as multi-class classification, includes a single class label for each instance. However, ophthalmic disease classification is a complex problem as multiple labels may be associated with a single instance. Among various publicly available fundus image datasets, only Ocular Disease Intelligent Recognition (ODIR-5K) dataset \cite{ref_11} presents the real-life challenge of multi-class multi-label ophthalmic disease detection. However, ODIR-5K is an imbalanced dataset. The biasing towards the majority class in an imbalanced dataset affects models' training and classification accuracy.

This work presents a DNN based framework for multi-label ophthalmic diseases classification of the fundus image. The proposed architecture improves multi-label classification accuracy by handling the issue of class imbalance in the fundus image dataset. The proposed dilated convolution based attention network named Discriminative Kernel Convolution Network (DKCNet) can simultaneously detect multiple lesion parts related to ophthalmic diseases appearing in a fundus image. Different backbone models are also used to evaluate their efficiency for multi-label classification. The proposed DKCNet achieves better performance for ophthalmic disease classification as compared to the methods proposed in the literature. 

\begin{figure*}[!t]
\centering
\includegraphics[width=1.0\textwidth]{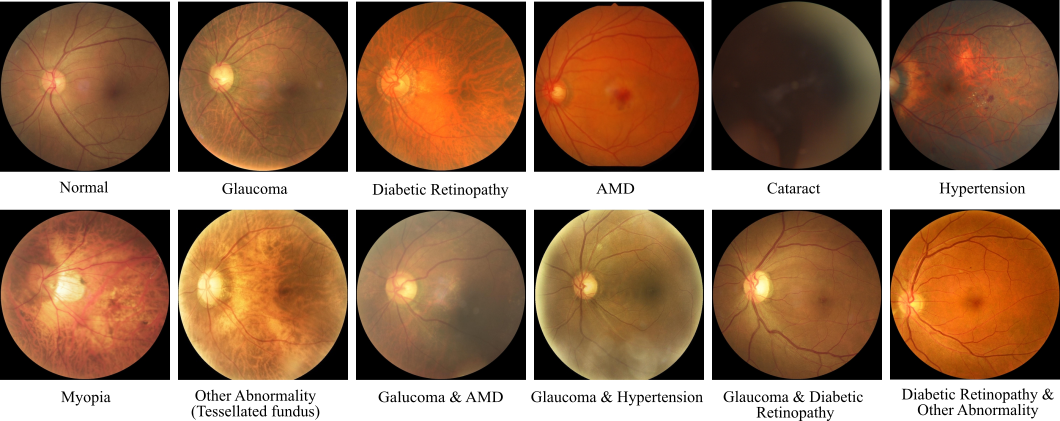}
\caption{Fundus images of different disease classes from ODIR-5K dataset \cite{ref_11}.}
\label{fig2}
\end{figure*}

The work is organized as follows. Section \ref{Diseases} connects retinal abnormalities in fundus images with ophthalmic diseases. State-of-the-art techniques for fundus image classification are discussed in Section \ref{related_works}. Section \ref{Methodology} describes the proposed transfer learning based DKCNet model for fundus image classification. The dataset used for experimental evaluation is also discussed here. The results are discussed in Section \ref{Results} and the work is concluded in Section \ref{Conclusion}.

\section{Ophthalmic Diseases Identified through Retinal Abnormalities in Fundus Images}
\label{Diseases}

Fig. \ref{fig2} shows fundus images belonging to different disease classes from the ODIR-5K dataset. Glaucoma is an eye condition that affects the optic nerve, whose strength is crucial for good vision. This harm is typically caused by an unusually high pressure in the eye \cite{ref_3}. Glaucoma can be distinguished by noticing changes in the proportion of the optic disc cup and neuro-retinal edge surface region known as the cup-to-disc ratio. Diabetic retinopathy is a diabetes intricacy that influences the tissues inside the retina. Diabetic retinopathy may not show any adverse indications, and patients may have minor issues related to clear vision. However, it may lead to visual impairment if not treated at an earlier stage. The macula is liable for clear central vision, and the distortion in vision starts if liquids collect in it. Age-related Macular Degeneration (AMD) can be distinguished by noticing the growth of fresh blood vessels or the presence of dead retinal cells known as neovascularization and geographic atrophy, respectively \cite{ref_5}. The cause of cataracts includes the development of patches that make vision difficult \cite{ref_4}. The presence of cataracts can be observed as optic disc, fovea, and other parts of the eye become hazy. Hypertension is a silent illness. This disease changes the biological shapes of veins, like length and thickness, and results in cardiovascular disease, stroke, and respiratory failures over the long run \cite{ref_6}. Myopia is a kind of eye issue that causes critical vision loss because of diminishing epithelial tissues and change in eye color. It fundamentally modifies the visualization of any object from a certain distance by making them blurred. The fundus images can also be used to detect some other kind of anomalies like pigment epithelium proliferation, the epiretinal membrane, tessellated fundus, and vitreous macular degeneration. 

\section{Related Works}
\label{related_works}
Deep learning in ocular imaging can be used in conjunction with telemedicine as a possible solution for selecting, diagnosing, and controlling ophthalmic diseases for patients in primary care \cite{ref_16}. Recent advances in neural network approaches are at the forefront of state-of-the-art disease recognition systems \cite{ref_17, ref_18}. Many researchers have made significant efforts to resolve the multi-label classification problem of ophthalmic disease. All these works are simulated on the publicly available ODIR-5K dataset. 

Islam et al. \cite{ref_12} proposed a shallow CNN-based model trained from scratch for classification of fundus images of ODIR-5K dataset. The left and right eye fundus images are input to the CNN model independently, and the disease label is assigned accordingly. Their approach made the disease classification model less complex, but their model is not able to distinguish multiple disease. Wang et al. \cite{ref_33} preprocessed fundus images using gray and color histogram equalization. Various data augmentation techniques are also used. The preprocessed gray and colored images are applied to two parallel EfficientNet models, and feature concatenation is done at the last layer for final classification. But they are able to achieve only 73\% AUC and 88 \% F1-Score on ODIR-5K dataset.

Li et al. \cite{ref_15} proposed a dense correlation network (DCNet) using transfer learning based ResNet architecture. The spatial correlation module (SCM) is the basic building block of this network architecture. The SCM block defines pixel-wise dense correlation between features extracted from color fundus images. These correlated features are fused to create the final feature map for classifying ophthalmic disease classes of ODIR-5K dataset with 93\% AUC and 91.3\% F1-score. Similarly, Gour and Khanna \cite{ref_13} proposed a pre-trained, two-input CNN architecture for the ODIR-5K dataset. They applied left and right eye fundus images to two parallel pre-trained VGG-16 simultaneously to extract the features \cite{ref_21}, which are concatenated to create a final feature map. In spite of the use of VGG model, they failed to beat the performance of \cite{ref_15}.

Li et al. \cite{ref_14} chose VGG-16, ResNet, Inception-v4, and Densenet \cite{ref_25} architectures with the sum, multiply, and concatenate operations on features extracted from the baseline model. They found that element-wise sum operation on feature maps yields better abnormality detection compared to the other two methods.  Lin et al. \cite{lin2021multi} proposed a graph convolution network (GCN) based self-supervising learning model known as MGC-Net. GCN is utilized to capture contextual information for multi-label fundus images whereas self-supervising learning is used for generalization of the network. In comparison to the backbone network, their model showed performance enhancement for fundus disease classification on ODIR dataset. Ou et al. \cite{ou2022bfenet} proposed two input CNN based attention model with multi-scale module for multi-label fundus image classification. Multi-scale module utilized $3\times3$ and $1\times1$ dilated convolution filters to capture multi-scale features. A spatial attention module is used for feature enhancement and learning inter-dependency between global and local information. The model is found computationally efficient but could not beat Li et al. \cite{ref_15} performance-wise.

ODIR-5k has a common target label for a pair of fundus image. Due to high variation in the sample counts of each class, it suffers from the class imbalance problem. None of the state-of-the-art methods attempt to address this issue. Class imbalance is a well-known issue in medical image classification, yet limited research is available on it in the context of deep learning methods \cite{ref_26}. Pratt et al. \cite{ref_27} implemented a CNN based model for the five-class classification of diabetic retinopathy disease. The authors demonstrated that the class-weight strategy can be used to resolve over-fitting and class imbalance issues. Buda et al. \cite{ref_26} examined the impact of class imbalance on classification problems using CNN. In their experimentation, CIFAR-10, MNIST, and ImageNet datasets are sub-sampled to construct artificially balanced datasets.

In CNN, the receptive field can be made large with increasing kernel size, but this usually results in an increasing number of learning parameters, which may lead to over-fitting problem \cite{Yu_Fisher}. To overcome this issue, Yu et al. \cite{Yu_Fisher} proposed dilated convolution operation which can enlarge the receptive field without adding additional computational cost. Since large receptive fields may not be able to recognise small objects, multi-scale feature extraction can be utilized to improve the image classification. Qi Zhang \cite{zhang2022novel} proposed a dilated convolution based network to extract multi-scale features with increased receptive field size. The model showed improved performance by extracting broader and deeper semantic information for liver tumour classification. Similarly, Tao et al. \cite{ku2021multilevel} proposed a multi-scale hybrid dilated convolution module for segmentation.  They used multiscale dilated convolution with variable dilation rate in the encoder-decoder architecture. Simulated on several backbone, CNN based model achieved improved performance for object segmentation. In this manuscript, a novel attention-based model is proposed with improved multi-label classification accuracy by resolving the class imbalance issue of the dataset.

\section{Methodology}
\label{Methodology}

\subsection{ODIR-5K Dataset}
ODIR-5K dataset used for experimentation in this work is made available online through a grand challenge by Peking University. The dataset contains around 5000 organized fundus images of the left and right eyes of patients. These fundus images were captured using different fundus cameras, like Kowa, Zeiss, and Cannon, having different image resolutions. Disease diagnostic keywords were assigned to these images from eye specialists \cite{ref_11}. Based on these diagnostic keywords, the disease classification labels are assigned to each pair of fundus images. This visual pathology dataset is unique in comparison to other publicly accessible data sets as it contains color fundus images of both left and right eyes of a patient with single/multiple abnormalities in a single image. The dataset contains a common target label for the pair of eye images. The images are grouped into eight disease classes, Normal, Diabetes, Glaucoma, Cataract, AMD, Hypertension, Myopia, and Others. Patients’ age and gender are also included. Another challenging part of the ODIR-5K dataset is that the other class images contain lesions related to 12 different ophthalmic diseases. It is not easy to learn appropriate features in such a case. Also, the dataset is highly imbalanced, considering the number of images in each of the eight classes. These issues negatively affect the accuracy and loss of the trained models for classification. Although ODIR-5K dataset is more applicable to real-life clinical situations as the images are captured with different cameras in different illumination conditions, this imposes a great challenge for any disease identification model.

\subsection{Pre-processing}
Most deep neural networks require dimension of input images in 1:1 aspect ratio. Images in the ODIR-5K dataset have different resolutions as these are captured from different cameras. Image crop operation is utilized to make these images appropriate for model training.  In image crop operation, the field of view is identified by seeking the start position of non-black pixel in the input fundus image. This position is used to identify the image mask border for crop operation. To support different DNN models, the image size needs to be adjusted explicitly. Therefore, the size of cropped image is kept to $224 \times 224$ pixels as it is commonly accepted size by most of the DNN models \cite{ref_21, ref_22, ref_23, ref_24, ref_25}.


\begin{table}[!h]
\caption{ODIR-5K dataset class statistics in original, after oversampling, and after undersampling operation.}
\label{table_3}
\centering
{\renewcommand{\arraystretch}{1.3}%
\begin{tabular}{|p{60pt}|p{25pt}|p{15pt}|p{30pt}|p{15pt}|p{30pt}|}
\hline
\multicolumn{1}{|c|}{\multirow{2}{*}{\textbf{Class (Label)}}} & \multirow{2}{*}{\textbf{Samples}} & \multicolumn{2}{c|}{\textbf{Oversampling}} & \multicolumn{2}{c|}{\textbf{Undersampling}} \\ \cline{3-6} 
\multicolumn{1}{|c|}{} & & \textbf{CBF} & \textbf{Samples} & \textbf{CBF} & \textbf{Samples} \\ \hline
Normal (N) & 1135 & 0 & 1135 & 12 & 95 \\ 
Diabetes (D)  & 1131 & 0 & 1131 & 11 & 103  \\ 
Glaucoma (G)  & 207 & 5 & 1035 & 2 & 104  \\ 
Cataract (C) & 211 & 5 & 1055 & 2 & 106  \\ 
AMD (AMD) & 171 & 7 & 1197 & 2 & 85 \\ 
Hypertension (H) & 94  & 12 & 1128 & 1 & 94 \\ 
Myopia (M) & 177 & 6 & 1062 & 2 & 86 \\ 
Others (O) & 944 & 0 & 944 & 10 & 95 \\ \hline
\end{tabular}}
\end{table}

\begin{figure*}[!t]
\centering
\includegraphics[width=1.0\textwidth]{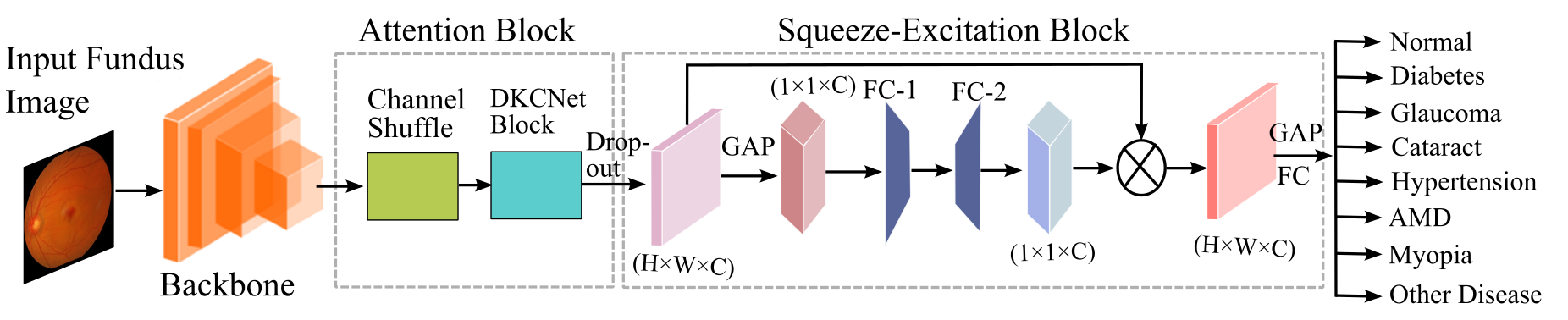}
\caption{Block diagram of the proposed DKCNet architecture.}
\label{fig3}
\end{figure*}

\begin{figure*}[!t]
\centering
\includegraphics[width=0.7\textwidth]{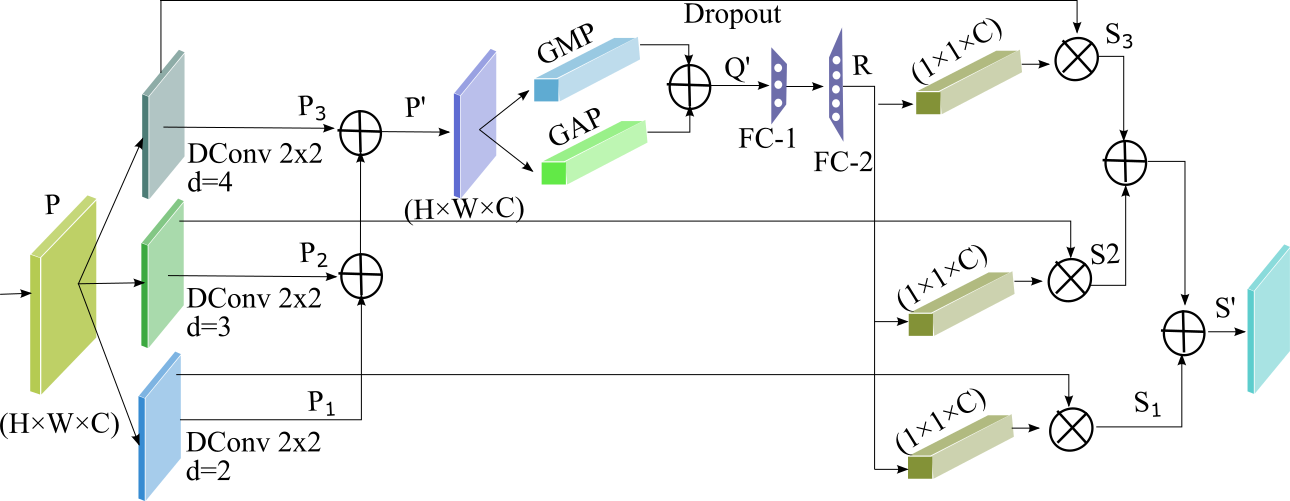}
\caption{Description of DKCNet Block.}
\label{fig4}
\end{figure*}

\subsection{Class Balancing}
The left and right fundus images are treated individually as input to the CNN architecture in this work. Based on the disease description appearing for a particular eye part, the disease label is being assigned to each image. A fundus image having multiple diseases is treated separately with individual disease class labels. The images with artifacts like ``low-quality image", ``optical disk photographically invisible," ``lens dust," and ``pimage offset" are removed from the final dataset to reduce the false recognition rate. After separating left and right eye images with their corresponding ground truth, the class-wise distribution is shown in Table \ref{table_3}. It is observed that three classes, normal, diabetes, and other classes have a significantly greater number of images in comparison to other disease classes. Random sampling techniques are used in this work to address the class balancing issue. Oversampling and undersampling are done to create two different versions of the datasets for implementation of the proposed CNN model.

\subsubsection{Oversampling} The most popular approach for producing synthetic image samples is to generate images with random attributes for minority classes. It can be seen in Table \ref{table_3}, classes Hypertension, Myopia, Cataract, AMD, and Glaucoma have a lesser number of samples compared to other classes. New samples for these classes are synthetically generated using different data augmentation strategies. The new sample size is calculated as per equation (\ref{eq:11}). 
\begin{equation}
\label{eq:11}
M_{minor}=N_{minor} \times (1+k)
\end{equation}
where $N_{minor} and M_{minor}$ represent the total number of samples in a minor class and synthetically generated samples for that class, respectively. Here, $k$ is defined as a class balancing factor (CBF) having a value ranging from 1 to 13, which represent the number of augmentation operations. For data augmentation, flip, re-scaling with scaling ratio (0.5, 0.7, 0.8 and 0.9), crop, rotation, contrast change, hue, saturation, and gamma value change operations are used.\\

\subsubsection{Undersampling} In this case, the whole dataset is shuffled and random images from majority classes have been selected to limit majority classes within the range of minority classes as per equation (\ref{eq:12}). 
 \begin{equation}
\label{eq:12}
M_{major}=\left \lfloor \frac {N_{major}}{k}\right \rfloor
\end{equation}

where $N_{major}, M_{major}$ represent the total number of samples in a major class and random samples for that class within a range, respectively. Again, $k$ is defined as class balancing factor (CBF). 

In this manner, image statistics for all 8 disease classes get equalized.

\subsection{Overview of the Proposed DKCNet Architecture}
As shown in Fig. \ref{fig3}, DKCNet is comprised of the backbone, attention block, and squeeze-excitation (SE) block. The backbone network is utilized to acquire the global feature maps. Any CNN based deep-learning model pre-trained on ImageNet dataset can be used as the backbone network to extract feature maps from the last layer of the model, which contains high-level semantic features of fundus image. These feature maps are $F_m \in R_m (H\times W\times C) $, where W, H, and C are width, height, and the number of channels in the feature maps. The output of the backbone network is fed into the attention block, which learns more region-wise features to discriminate lesion parts. A dropout layer follows this to reduce overfitting with a drop rate of 0.3. These discriminative features are processed by SE block, which dynamically re-calibrate channels. Finally, a global average pooling layer followed by a fully connected layer performs disease classification by predicting class label probability.

\subsection{DKCNet Block}
The standard convolution with a fixed kernel captures contextual information by sliding on the feature maps. In this case, features of a similar group of pixels may have a different representation in other regions, resulting in intra-class inconsistency. It is widely accepted that more contextual information can be captured by generating multi-scale features using different receptive field sizes \cite{ref_29, ref_30}. Dilated convolution captures multi-scale information by varying kernel sizes known as dilation rate. With a large receptive field size, more semantic information can be captured. As depicted in fig. \ref{fig4}, dilated convolution is utilized in the proposed model to capture multi-scale features without adding extra computational cost. 

The extracted feature maps from the backbone are passed to the channel shuffle block, which permutates the channel and permits data stream across feature channels. The DKCNet Block takes input from the channel shuffle block. It applies dilated convolution operation using $ 2 \times 2 $ kernel with dilation rate ranging from 1 to 3. Dilated convolution (DConv) is followed by batch normalization ($f_{BN}$) and ReLu activation ($f_{ReLu}$) function as shown in equation (\ref{eq:1}).

\begin{equation}
\label{eq:1}
P_i = f_{ReLu}(f_{BN}(DConv_d(P))), i=\left \{1,2,3\right\}, d=\left \{2,3,4\right\}
\end{equation}

Features obtained after dilated convolution operation are grouped together by element wise addition operation as per equation (\ref{eq:2}). 
\begin{equation}
\label{eq:2}
P' = P_1 \oplus P_2 \oplus P_3
\end{equation}

Now the spatial information is squeezed from the feature maps by performing global average pooling ($f_{GAP}$) and global max-pooling ($f_{GMP}$) operations followed by element-wise addition to get global spatial information as per equation (\ref{eq:3}).
\begin{equation}
\label{eq:3}
Q = f_{GMP}(P') \oplus f_{GAP}(P')
\end{equation}

This squeezed spatial information is passed through two fully connected layers for channel dimension reduction by a factor $r$. Some of the features are then dropped by introducing a feature drop layer with dropout rate of 0.25 followed by sigmoid ($\sigma$) activation function as per equation (\ref{eq:4}) \& (\ref{eq:5}).
\begin{equation}
\label{eq:4}
Q' = f_{Drop=0.25}(Q)
\end{equation}

\begin{equation}
\label{eq:5}
R = f_{Sigmoid}(f_{FC}(Q'))
\end{equation}

After that, obtained squeezed information vector is element-wise multiplied with feature maps obtained by dilated convolution as shown in equation (\ref{eq:6}).
\begin{equation}
\label{eq:6}
S_i = R \otimes P_i, i=\left \{1,2,3\right\}
\end{equation}

\begin{equation}
\label{eq:7}
S' = S_1 \oplus S_2 \oplus S_3
\end{equation}

Finally, as per equation (\ref{eq:7}), the obtained features are added together to get the attention map.

\subsection{Loss Function}
This work deals with a multi-class multi-label classification problem; one or more disease labels are required as the output for each left and right eye image input. For the computation of the difference between predicted target labels and actual labels, the Binary Cross-Entropy (BCE) loss function is used, which is given as:

\begin{equation}
\label{eq:8}
BCE( y, \hat{y}) = - \frac{1}{M}\sum_{i=1}^{M}y_i \log (\hat{y}) + (1-y_i)\log(1-\hat{y})
\end{equation}

where ${M}$ is the number of samples in the training set, ${y}$ is the actual label, and $\hat{y}$ is the predicted label.

\subsection{Experimentation Setup}
In this study, a backbone network is selected via experimentation with pre-trained ResNet, InceptionV3, and InceptionResNet architectures. The ODIR-5K dataset is split into 80\% and 20\% for the training and validation set, respectively. For oversampling, CBF value is selected as 12, 6, 5, 7, 5 for hypertension, myopia, cataract, AMD and glaucoma disease classes, respectively; and 0 for normal, diabetes, and other disease classes. Similarly, for under-sampling, CBF value is selected as 12, 11, 10, 1 for normal, diabetes, others, and hypertension classes; and 2 for glaucoma, cataract, AMD, and myopia class. Samples synthetically generated by both oversampling and undersampling are depicted in Table \ref{table_3}. Model is optimized with SGD optimizer. The initial learning rate is set to 0.0005 with decay factor $1e^{-6}$ along with the batch size as 16. All the models are trained for 100 epochs with BCE loss function. All experimentation work is performed on NVIDIA T4 GPU with 16 GB memory. Two experimentation scenarios have been implemented to investigate the performance of DKCNet. In the first scenario, the model is trained with a backbone network for the classification of eight ophthalmic diseases. In the second scenario, training is done on the fusion of the backbone with DKCNet.

\begin{figure*}[!t]
\centering
\includegraphics[width=0.9\textwidth]{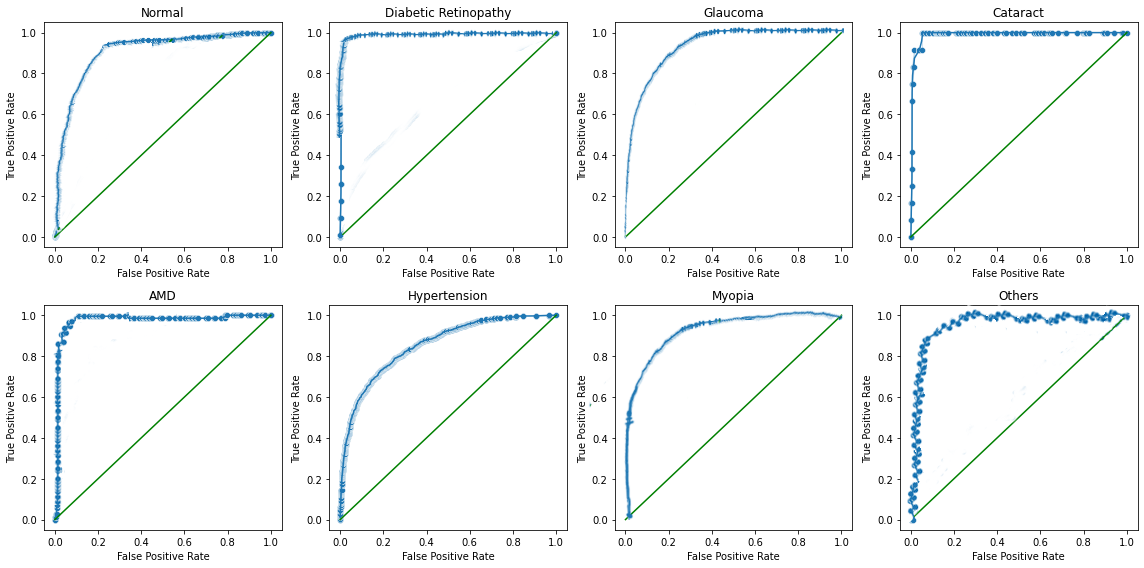}
\caption{AUC curves obtained with the proposed model for multi-class multi-label classification of retinal disease.}
\label{fig7}
\end{figure*}
 
\begin{figure*}[!t]
\centering
\includegraphics[width=0.55\textwidth]{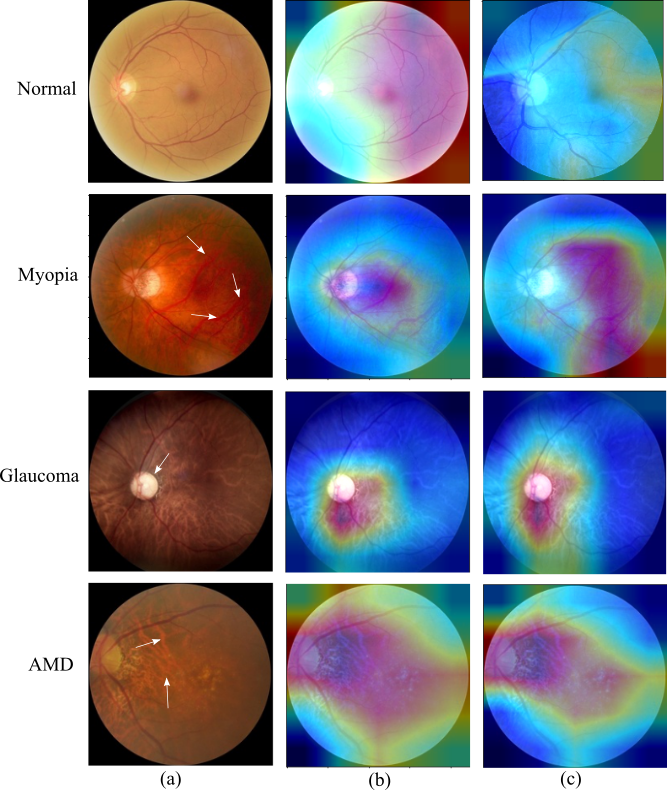}
\caption{Class activation maps obtained with DKCNet for single diseased fundus images from ODIR-5K dataset (a) original fundus images with lesion parts highlighted using white arrow, (b) heatmaps generated by backbone network, and (c) heatmaps refined by DKCNet.}
\label{fig5}
\end{figure*}

\begin{figure*}[!t]
\centering
\includegraphics[width=0.9\textwidth]{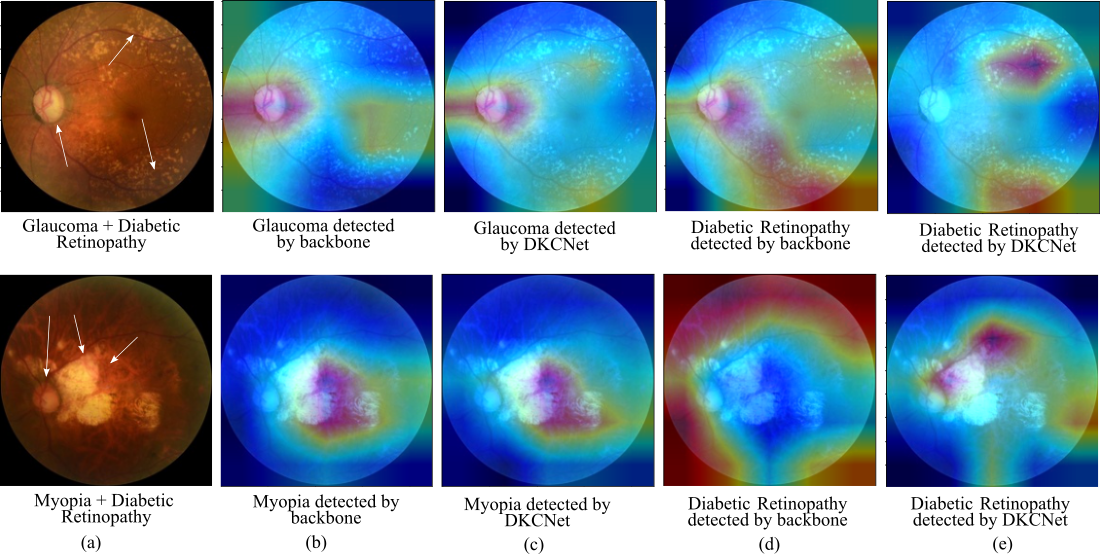}
\caption{Class activation maps obtained with DKCNet for multiple diseased fundus images from ODIR-5K dataset (a) original fundus images with lesion parts highlighted using white arrows, (b) heatmaps generated by backbone network, and (c) heatmaps refined by DKCNet.}
\label{fig6}
\end{figure*}

\section{Results \& Discussion}
\label{Results}
The performance of the proposed model is evaluated using Area Under receiver operating Curve (AUC), F1-Score, and kappa score. AUC curve is a performance estimator for classification problems. The AUC value closer to 1 means better performance of the model to classify the disease labels. Similarly, F1-score is defined as the harmonic mean of recall and precision values as shown in equation (\ref{eq:15}). Kappa score is a proportion of how intently the instance classified results matched with the ground truth label. It is calculated as per equation (\ref{eq:16}).

\begin{equation}
\label{eq:13}
Precision = \frac{T_p}{F_p+T_p}
\end{equation}

\begin{equation}
\label{eq:14}
Recall = \frac{T_p}{F_n+T_p}
\end{equation}

\begin{equation}
\label{eq:15}
F1 = 2 \times\frac{Precision \times Recall}{Precision + Recall}
\end{equation}

\begin{equation}
\label{eq:16}
k = \frac{(p_o - p_e)}{1 - p_e}
\end{equation}

where ${T_p, F_p, F_n}$ are true positive, false positive, and false negative, respectively; $p_o$ is the empirical probability of agreement on the label assigned to the sample and $p_e$ is the expected agreement when both annotators assign labels randomly.

The effectiveness of DKCNet is experimentally investigated, and results for 10-fold cross validation are shown in Table \ref{table_4}. The proposed method can be applied to a wide range of backbone networks to improve ophthalmic disease classification. ResNet-101, Inception V3, and InceptionResNet backbone networks are implemented with the three different baselines (without sampling, oversampling and undersampling) in the proposed approach. The increase in performance can be seen in both over-sampled and under-sampled datasets compared to without sampling. Results reported in Table \ref{table_4} indicate that the backbone network integrated with DKCNet achieves a significant performance improvement (96.08 AUC, 94.28 F1-Score, and 0.81 kappa-score) for ophthalmic disease classification with over-sampling of minority classes for training. For each test image, we assigned target labels with a confidence greater than 0.5 to be positive, and compared them with a ground truth labels. Fig \ref{fig7} shows the AUC curves for each disease class classification performance.


\begin{table}[!t]
\caption{Ophthalmic disease classification by using DKC block with state-of-the-art backbone networks on ODIR-5K dataset.}
\label{table_4}

{\renewcommand{\arraystretch}{1.3}%
\begin{tabular}{|p{48pt}|p{16pt}|p{20pt}|p{14pt}|p{18pt}|p{14pt}|p{18pt}|}
\hline
\multicolumn{1}{|c|}{\multirow{2}{*}{\textbf{Model}}} & \multicolumn{2}{c|}{\textbf{No Sampling}} & \multicolumn{2}{c|}{\textbf{Undersampling}} & \multicolumn{2}{c|}{\textbf{Oversampling}} \\ \cline{2-7} 
\multicolumn{1}{|c|}{} & \multicolumn{1}{l|}{\textbf{AUC}} & \textbf{F1-Score} & \multicolumn{1}{l|}{\textbf{AUC}} & \textbf{F1-Score} & \multicolumn{1}{l|}{\textbf{AUC}}  & \textbf{F1-Score} \\ \hline
ResNet-101 & \multicolumn{1}{l|}{70.12} & 78.53 & \multicolumn{1}{l|}{87.04} & 89.56 & \multicolumn{1}{l|}{94.33} & 92.28 \\ \hline
ResNet-101+ DKC Block  & \multicolumn{1}{l|}{70.26} & 79.28 & \multicolumn{1}{l|}{86.53} & 88.15 & \multicolumn{1}{l|}{93.52}  & 91.62  \\ \hline
InceptionV3 & \multicolumn{1}{l|}{72.94} & 78.17 & \multicolumn{1}{l|}{87.45} & 88.93 & \multicolumn{1}{l|}{86.31} & 87.37  \\ \hline
InceptionV3+ DKC Block & \multicolumn{1}{l|}{73.86} & 79.44 & \multicolumn{1}{l|}{\textbf{88.24}} & \textbf{88.93} & \multicolumn{1}{l|}{88.59}  & 89  \\ \hline
InceptionResNet & \multicolumn{1}{l|}{74.22} & 80.44 & \multicolumn{1}{l|}{86.94} & 88.68 & \multicolumn{1}{l|}{94.24}  & 91.53 \\ \hline
InceptionResNet + DKC Block & \multicolumn{1}{l|}{\textbf{74.55}} & \textbf{80.96} & \multicolumn{1}{l|}{88.05} & 88.93 & \multicolumn{1}{l|}{\textbf{95.4}} & \textbf{93.18} \\ \hline
\end{tabular}}
\end{table}

\begin{table}[!t]
\caption{Ophthalmic disease classification results obtained with different methods on ODIR-5K dataset.}
\label{table_5}
\setlength{\tabcolsep}{3pt}
{\renewcommand{\arraystretch}{1.3}%
\begin{tabular}{|p{65pt}|p{60pt}|p{20pt}|p{20pt}|p{25pt}|p{20pt}|}
\hline
\textbf{Author} & \textbf{Method} & \textbf{AUC} & \textbf{F1-Score} & \textbf{Params (M)} & \textbf{Flops (G)} \\ \hline
Islam et al. \cite{ref_12}, 2019 & Shallow CNN & 80.5 & 85 & 1.1 & -\\
Wang et al. \cite{ref_33}, 2020 & EfficientNet & 73 & 88 & - & - \\
Gour and Khanna \cite{ref_13}, 2020 & Two Input VGG-16 & 84.93 & 85.57 & 15.2 & 80.2 \\
Li et al. \cite{ref_15}, 2020 & ResNet-101 & 93 & 91.3 & 74.2 & 68.7 \\
Ning Li et. al. \cite{ref_14}, 2021 & Inception-v4 & 88 & 85.93 & - & - \\
Lin et. al. \cite{lin2021multi}, 2022 & Graph Conv. Network & 78.16 & 89.66 & - & - \\
Ou et. al. \cite{ou2022bfenet}, 2022 & ResNet-50 & 90.3 & 88.6 & 82.6 & 67 \\
\textbf{Proposed Method} & \textbf{InceptionResnet + DKC Block} & \textbf{96.08} & \textbf{94.28} & \textbf{87.7} & \textbf{13.4} \\ \hline
\end{tabular}}
\end{table}

The proposed model is also compared with five recent multi-class, multi-label ophthalmic disease classification models to show its efficacy. The results are reported in Table \ref{table_5}. As mentioned in Section \ref{related_works}, Islam et al. \cite{ref_12} considered single shallow CNN model which could not perform well for multi-label classification. Wang et al. \cite{ref_33} obtained a better F1-score with EfficientNet model but lacked in AUC compared to other methods. Furthermore, Gour and Khanna \cite{ref_13} achieved slightly better performance. But they utilized a heavy VGG16 model, which does not contain a batch normalization layer, making it hard to converge. The performance obtained by Ning Li et. al. \cite{ref_14} by using inception-v4 model with element-wise sum feature fusion is comparable with that obtained by Gour and Khanna \cite{ref_13}. The spatial corelation model proposed by Li et al. \cite{ref_15} delivered better outcomes for different mixes of ResNet structures. The model performs better among counterpart models in terms of F1-score and AUC. However, the proposed DKCNet model achieves improvement in AUC and F1-Score by 2.5\% and 2.05\%, respectively as compared to those achieved by Li et al. \cite{ref_15}. Table \ref{table_5} also shows that the proposed network requires less number of floating-point operations per second (FLOPS) compared to state-of-the-art methods.

\begin{table}[!t]
\caption{Cross dataset performance evaluation of the proposed model.}
\label{table_6}
\setlength{\tabcolsep}{3pt}
{\renewcommand{\arraystretch}{1.3}%
\begin{tabular}{|p{170pt}|p{25pt}|p{35pt}|}
\hline
\textbf{Testing Dataset} & \textbf{AUC} &  \textbf{F1-Score} \\ \hline
Messidor (Diabetic Retinopathy)  & 89.37 & 87.75  \\ \hline
G1020 (Glaucoma)  & 93.14 & 91.42  \\ \hline
Joint Shantou International Eye Centre (Multi class) & 94.18 & 91.15 \\ \hline
\end{tabular}}
\end{table}

To check the biasness of proposed model towards training data, it is further tested on three publicly available benchmark datasets: Messidor (Diabetic Retinopathy), G1020 (Glaucoma), Joint Shantou International Eye Centre (Multi class). From the Table \ref{table_6}, it can be observed that the proposed DKCNet can predict retinal diseases effectively on completely unseen fundus image datasets. 



Qualitative analysis of results is performed by visualizing activation maps using Grad-CAM \cite{ref_28}. In Fig. \ref{fig5}, the first column shows fundus images containing a single disease class with lesion part marked with a white arrow. The results of the backbone network, i.e., InceptionResnet are visualized in the second column. The third column shows refined activation maps obtained by using DKCNet with the backbone network. In the first row, the results generated by the backbone network show false detection of lesion part in a normal eye image, whereas the proposed model is able to discriminate such situations effectively. Similarly, the backbone network failed to detect some lesion parts in the input image in the second row, but the proposed model can identify those efficiently. The third row corresponds to the cases where the prediction results obtained with the backbone network and the proposed model are similar. In the fourth row, the backbone network falsely highlights the lesion part in the larger portion of the eye, whereas the proposed model refines the detection and is near the ground truth.

Similarly, multi-class classification performance can be visualized in Fig \ref{fig6}. Here, the first column shows input fundus images of patients' eyes having multiple diseases. The second and third columns correspond to class activation maps generated by the backbone network for predicted class, whereas the last two columns demonstrate a refined class activation map obtained with DKCNet with better visual classification.  In the fourth column of the first row of Fig \ref{fig6}, it can be seen that the backbone network failed to detect some of the lesion parts, whereas the proposed method detects those well. In the second row, the backbone network shows false detection for lesion parts, whereas the proposed DKCNet highlights those parts more accurately, as seen in the fourth and the last column of Fig \ref{fig6}.

\section{Conclusion and Future Work}
\label{Conclusion}
The DKCNet architecture proposed in this work enables CNN based model to learn discriminative features with an attention module without introducing extra cost. The novelty of this work lies in a model which improves ophthalmic disease classification performance and solves the class balancing issue of the highly imbalanced ODIR-5K dataset having multiple common labels for fundus image pair of a patient’s left and right eye. DKCNet is composed of an attention block followed by a SE block. The attention block takes features from the backbone network and generates discriminative feature attention maps. The SE block takes the discriminative feature maps and performs channel wise attention almost at no computational cost. The experimentation has been done with three backbone CNN-based architectures, and it is found that the proposed DKCNet shows superior performance compared to counterpart methods with InceptionResnet model.  As it is expensive to obtain good quality and labeled fundus images, it is planned to use generative adversarial networks to generate samples for minority classes artificially. Also, the model can be explored to localize and find the types of lesions.

\bibliographystyle{IEEEtran}
\bibliography{bibliography}

\end{document}